# The Structure of Integrable One-Dimensional Systems

Bill Sutherland
Department of Physics
University of Utah
Salt Lake City, UT 84112
vsuther@comcast.net
August 1, 2007

Abstract: We explain the relationship between the classical description of an integrable system in terms of invariant tori and action-angle variables, and the quantum description in terms of the asymptotic Bethe ansatz.

§1  We wish to describe the structure of integrable one-dimensional many-body systems. We discuss both the classical and the quantum regimes, and especially the connection between the two. Although we do not give proofs for our assertions, we believe that the picture we present is both internally consistent and consistent with the large body of disjoint results existing in the literature, and thus is compelling. We begin our discussion with a system given by the Hamiltonian

$$H = \frac{1}{2m}\sum_j p_j^2 + \sum_{j>i}\sum U(x_j - x_i). \tag{1}$$

For simplicity, we will here assume that the pair-potential is symmetric, repulsive, impenetrable, with a force decreasing to zero sufficiently rapidly. We can thus assume that the particles are ordered, so $x_j > x_i$ for $j > i$. We assume the potential is such that the system is integrable. A non-trivial example of such a potential is

$$U(r) = \frac{g}{\sinh^2 r}. \tag{2}$$

The effect of statistics is minimal for impenetrable potentials in one dimension; we shall here assume that the particles are identical bosons. (All these restrictions can later be relaxed somewhat.)

Although we have assumed an infinite line where $+\infty > j > -\infty$, in fact, to make sense of such an infinite system we only allow solutions which are *periodic in the index*, so that $x_{j+N} = x_j + L$ for some integer $N > 0$ and $L > 0$; we refer to these as *finite* solutions. These solutions then allow us to take the *thermodynamic limit* $N, L \to \infty$ with *density* $d \equiv N/L \equiv 1/a$ finite. This is equivalent to the finite, periodic Hamiltonian

$$H[N,L] = \frac{1}{2m}\sum_{j=1}^{N} p_j^2 + \sum_{j>i=1}^{N} U(x_j - x_i \mid L), \qquad (3)$$

where

$$U(x \mid L) \equiv \sum_{j=-\infty}^{\infty} U(x - jL) \underset{L \to \infty}{\to} U(x) \qquad (4)$$

is the *periodic potential*. Thus, $H[N, L]$ is the energy per period of the infinite $H$ of eq. (1), less the energy of interaction of a particle with its periodic images, given by

$$\frac{1}{2} \sum_{\substack{j=-\infty \\ (\neq 0)}}^{\infty} U(jL) \underset{L \to \infty}{\to} 0. \qquad (5)$$

§2   There is a well established theory of integrable systems in the classical case. [1,2] If our finite system of $N$ particles is integrable, then there exist $N$ independent integrals of motion, in involution. This implies that the trajectory is confined to an $N$-dimensional manifold, and that this manifold is covered by $N$ compatible vector fields – the flows determined by the vector fields. Then, by a theorem of topology, this manifold has the shape of an $N$-dimensional torus. That is, the manifold can be deformed smoothly into an $N$-dimensional cube with opposite sides identified, and this deformation allows us to define new coordinates – the *angle variables* $\vartheta = \{\vartheta_1, \cdots, \vartheta_N\}$ – which vary from 0 to $2\pi$. If the $N$ compatible vector fields are taken as a basis for these coordinates, then the

vector field $(-\partial H/\partial p, \partial H/\partial x)$ determining the flow of the Hamiltonian, has constant components $\omega = \{\omega_1, \cdots, \omega_N\}$ throughout the manifold. That is, the trajectories become straight lines $\vartheta(t) = \omega t + \vartheta_0$.

We take the angle variables as a new set of position coordinates, and then construct the canonically conjugate momenta $A = \{A_1, \cdots, A_N\}$, called the *action variables*. Together, we have the *action-angle variables* $(A, \vartheta)$. The action variables can be determined from the invariance of the *action function*

$$S(x'',x',E) = \int_{x'}^{x''} pdx = \int_{\vartheta'}^{\vartheta''} Ad\vartheta. \tag{6}$$

If only a single $\vartheta_i$ varies from 0 to $2\pi$, this defines a closed curve $C_i$ in $(p,x)$ space, and we find that

$$A_i = \frac{1}{2\pi} \oint_{C_i} pdx. \tag{7}$$

Upon making the canonical transformation from $(p,x)$ to $(A,\vartheta)$, the new Hamiltonian $H[A]$ does not depend on the angle variables, and the equations of motion become

$$\begin{aligned} \frac{dA}{dt} &= -\frac{\partial H}{\partial \vartheta} = 0, \\ \frac{d\vartheta}{dt} &= \frac{\partial H}{\partial A} = \omega. \end{aligned} \tag{8}$$

Again, the complete solution is $\vartheta(t) = \omega t + \vartheta_0$.

§3   For integrable systems such as ours, which support scattering, we have argued that the asymptotic Bethe ansatz gives a complete description of the spectrum of the Hamiltonian in the quantum case. [3,4] Because of the conservation laws, when $N$ particles scatter, the out-going asymptotic momenta are confined to be permutations of the in-coming momenta. Thus, the asymptotic wavefunction is of the form

$$\Psi(x) \to \sum_P \Psi(P) \exp[i \sum_{j=1}^{N} x_j k_{Pj}], \tag{9}$$

where $P$ is one of the $N!$ permutations of the incoming momenta $k = (k_1, \cdots, k_N)$. (Although we call these $k$'s 'asymptotic momenta', they are really 'asymptotic wave numbers', since the true asymptotic momenta are given by $\hbar k$. This should cause no confusion, since it is really only a difference in the choice of units.) Further, the amplitudes $\Psi(P)$ must be related by two-body scattering, so

$$\Psi(P')/\Psi(P) = -\exp[-i\theta(k-k')], \tag{10}$$

where $\theta(k)$ is the two-body phase shift, and the permutations $P, P'$ are the same, except $Pj = P'(j+1), P(j+1) = P'j$, so that $k \equiv k_{Pj}, k' \equiv k_{P'j}$.

Upon imposing periodic boundary conditions, the total phase change when a particle is taken around the ring of circumference $L$ must be a multiple of $2\pi$. Thus we have a set of $N$ equations determining the asymptotic momenta

$$k_j L = 2\pi I_j + \sum_{\substack{i=1 \\ (i \neq j)}}^{N} \theta(k_j - k_i), \tag{11}$$

where the $I_j$'s are distinct quantum numbers equal to integers (half-odd integers) for an odd (even) number of bosons. We assume that they are ordered so that $I_N > \cdots > I_j > \cdots > I_1$. We call these $I$'s *hard-core boson* quantum numbers; for $N$ odd, they are the same as free fermion quantum numbers. The total momentum $P$ and energy $E$ are given by

$$P = \hbar \sum_{j=1}^{N} k_j,$$
$$E = \frac{\hbar^2}{2m} \sum_{j=1}^{N} k_j^2. \tag{12}$$

We could as well use *free boson* quantum numbers $J_N \geq \cdots \geq J_j \geq \cdots \geq J_1$ with $J$ integer, by simply redefining

$$J_j \equiv I_j - \sum_{\substack{i=1 \\ (i \neq j)}}^{N} \text{Sign}[j-i],$$

$$\theta^B(k) \equiv \theta(k) + \text{Sign}[k], \quad , \tag{13}$$

$$k_j L = 2\pi J_j + \sum_{\substack{i=1 \\ (i \neq j)}}^{N} \theta^B(k_j - k_i).$$

In particular, the ground state is given by choosing the quantum numbers to be densely spaced, so that in the thermodynamic limit the $k$'s are distributed with a density $\rho(k)$, where $L\rho(k)dk$ gives the number of $k$'s between $k \to k + dk$. Thus, if we differentiate eq. (11), we obtain the integral equation

$$1 = 2\pi\rho(k) + \int_{k_1}^{k_2} \theta'(k-k')\rho(k')dk', \quad k_2 \geq k \geq k_1. \tag{14}$$

Upon solving this integral equation, we can then evaluate parametrically

$$N/L = d = \int_{k_1}^{k_2} \rho(k)dk,$$

$$P/L = p = \hbar \int_{k_1}^{k_2} k\rho(k)dk, \tag{15}$$

$$E_0/L = e_0[p,d] = \frac{\hbar^2}{2m} \int_{k_1}^{k_2} k^2\rho(k)dk \equiv e_0[d] + \frac{p^2}{2md}.$$

Since we often reference low-lying states to the ground state, let us go to the center of mass frame $P = p = 0$ and take as the *standard form* for the equations determining the ground state of the system at rest,

$$1 = 2\pi\rho(k) + \int_{-q}^{q} \theta'(k-k')\rho(k')dk', \quad |k| \leq q,$$

$$d = \int_{-q}^{q} \rho(k)dk, \tag{16}$$

$$e_0 = \frac{\hbar^2}{2m} \int_{-q}^{q} k^2 \rho(k)dk$$

This gives the zero temperature thermodynamics $e_0[d]$ when we eliminate the parameter $q$.

§4   We now answer the question: What is the relationship between the classical description in terms of invariant tori and action-angle variables, and the quantum description in terms of the asymptotic Bethe ansatz and integral equations for the distribution of asymptotic momenta?

§5   We begin with a semiclassical description of free motion on the invariant tori. The action variables are quantized in steps of $\hbar$. However, the action operators correspond to classical action variables, so we can only conclude that $A_j = \hbar(M_j + \beta_j)$. There are $N-1$ bound or *collective* degrees of freedom – essentially quantized density waves, or beams of interacting phonons – which we label as $j = 1, \cdots, N-1$. Since the action variables are adiabatic invariants, they are unchanged when we turn off the interaction to give free particles, or when we take the semiclassical limit $\hbar \to 0$ by increasing the interaction to infinity. In the semiclassical limit, we make the harmonic approximation and conclude

$$A_j = \hbar(M_j + 1/2), \text{ with } M_j = 0, 1, 2, \cdots, \text{ for } j = 1, \cdots, N-1. \qquad (17)$$

In addition we have the center of mass degree of freedom which we identify by $j = 0$. The center of mass momentum – or total momentum – is given by

$$P = 2\pi\hbar N(M_0 + \beta_0)/L = 2\pi\hbar d A_0, \qquad (18)$$

with $M_0 = \cdots, -1, 0, 1, 2, \cdots$. Because of the periodic boundary conditions, $N\beta_0$ must be an integer. If we begin with an eigenstate of momentum $P$, then we can only boost $P$ by integer multiples of $2\pi\hbar d$. What is important to note is that because of the term $\beta_0$, in general we cannot begin with momentum $P$, and boost to the center of mass frame with $P = 0$. Instead, the best we can do is to 'umklapp' by a discrete Galilean transformation into the *Brillouin zone* with a range of momentum $\Delta P = 2\pi \hbar d$. This situation is quite different from the classical system.

From the limit of free particles, we find

$$P = 2\pi \hbar d[M_0 + \sum_{j=1}^{N-1} jM_j / N] \equiv 2\pi \hbar d M_0 + \hbar \sum_{j=1}^{N-1} \kappa_j M_j. \qquad (19)$$

Thus, we identify

$$N\beta_0 = \sum_{j=1}^{N-1} jM_j \bmod N, \qquad (20)$$

with the Brillouin zone $2\pi \hbar d > P \geq 0$ corresponding to $M_0 = 0$. Occasionally we will shift the Brillouin zone; for instance, eq. (19) seems to imply that the phonons carry momentum $\hbar \kappa$, and so it is useful to have the zone shifted so that $\pi d \geq \kappa > -\pi d$.

Let us begin with a state $\{M_0, M_1, \cdots, M_{N-1}\}$, with momentum $P$ and energy $E$; we eliminate $M_0$, write $M \equiv \{M_1, \cdots, M_{N-1}\}$, and express the energy as $E[P, M]$. We <u>define</u> the center of mass energy as

$$E[M] \equiv E[P, M] - \frac{P^2}{2mN}. \qquad (21)$$

However, we cannot actually boost to this energy; the best we can do is boost to the Brillouin zone $P \to P_0$, with energy

$$E[P, M] \to E[P_0, M] = E[M] + \frac{P_0^2}{2mN}. \qquad (22)$$

The last term in the energy is only of order $1/L$, while $P_0$ is of order 1, so in many situations – but not all – this extra term is negligible.

The absolute ground state is given by all $M_j = 0$, and has energy $E_0 \equiv E[0]$ and momentum $P = P_0 = 0$. In one dimension, zero point fluctuations are severe enough in to destroy the long-ranged crystalline order of the classical equilibrium state.

The lowest energy states are of order $1/L$ above the ground state, and momentum and energy are of the form

$$P = \frac{2\pi\hbar}{L}\sum jM_j \equiv \hbar\sum \kappa_j M_j,$$
$$E[M] - E_0 = \hbar\sum \omega_j M_j = \frac{2\pi\hbar v_s}{L}\sum |j|M_j = \hbar v_s \sum |\kappa_j|M_j. \quad (23)$$

Here we have shifted the Brillouin zone, so that the summation is over positive and negative integers $j$ such that $\sum |j|M_j \ll N$. The velocity of sound is $v_s$. These states constitute the *conformal cone*, and determine the long ranged correlations in the ground state.

§6  We now consider states with all collective degrees of freedom frozen out except for $G = 0, 1, 2, \cdots$ of them, the ground state being $G = 0$. Classically, such a $G$-subspace corresponds to invariant tori of *genus* $G + 1 \equiv \Gamma$ when we include the center of mass motion. In the quantum case, this $G$-subspace of states is spanned by eigenstates with all action quantum numbers zero except for a subset $G$ of them, given by $M_\alpha = 1, 2, \cdots$, with $\alpha = 1, \cdots, G$. This index $\alpha$ is no longer the index of all action variables as in §5, but instead indexes only those non-zero action variables which are quantized in other than the lowest eigenstate. We will specify (indirectly) which collective modes are excited, by choosing a set of $G + 1$ positive integers $N_\alpha = 1, 2, \cdots$, $\alpha = 0, 1, 2, \cdots, G$. (The relation between these $N_\alpha$ and the previous index $K_\alpha$ for the collective modes will be explained shortly.) These integers are not independent, since we require that they sum to $N$. Finally, we specify the total momentum by the integer quantum number $M_0 = \cdots, -1, 0, 1, 2, \cdots$. Thus our states are labeled by $2(G + 1) = 2\Gamma$ integer quantum numbers $[M, N] \equiv [\{M_0, M_1, \cdots, M_G\}, \{N_0, N_1, \cdots, N_G\}]$.

Using adiabatic arguments, let us describe how such a state looks in three different schemes.

First, if we take turn off the interaction, we recover the limit of *hard core bosons or free fermions*. We find that the quantum numbers are distributed in $G + 1$ dense blocks of $N_\alpha$ occupied orbitals, separated by $G$ dense blocks of $M_\alpha$, $\alpha \neq 0$ unoccupied orbitals; the two alternate in order. This is the *fermion scheme*.

Second, by an appropriate limit (see §3) we can also recover free bosons. In this case exactly $G+1$ orbitals are occupied with occupation numbers $N_\alpha$; the occupied orbitals are again separated by $G$ dense blocks of $M_\alpha$, $\alpha \neq 0$ unoccupied orbitals, with the two alternating in order. This is the *boson scheme*.

Finally, if we take the semiclassical limit, we find that all collective modes are frozen with $M = 0$, except for $G$ of them. The action quantum numbers for these excited modes are $M_\alpha$. As previously discussed in §5, we label the collective modes by an integer $K = 1, 2, \cdots, N-1$, and the labels $K_\alpha = 1, 2, \cdots$, $\alpha = 1, \cdots, G$ for these excited modes are

$$
\begin{aligned}
K_1 &= N - N_0 = N_1 + \cdots + N_G, \\
K_2 &= N - N_0 - N_1 = N_2 + \cdots + N_G, \\
&\vdots \\
K_\alpha &= N - \sum_{\beta=1}^{\alpha-1} N_\beta = \sum_{\beta=\alpha}^{G} N_\beta, \\
&\vdots \\
K_G &= N_G.
\end{aligned}
\quad (24)
$$

It is useful to define $K_0 = N$. We see that $N = K_0 > K_1 > K_2 > \cdots > K_G > 0$. The $K$'s are as good quantum numbers as the $N$'s; we write them collectively as $K = \{K_1, \cdots, K_G\}$. This is the *phonon scheme*, and best fits the semiclassical picture of action-angle variables and motion on invariant tori.

In figure (1) we show the three schemes for a simple example with $N = 11$. The total momentum is given as

$$
P = \frac{2\pi\hbar}{L} \left[ NM_0 + \sum_{\alpha=1}^{G} M_\alpha K_\alpha \right].
\quad (25)
$$

Thus, the number of collective degrees of freedom $G$ acts as a conserved number of particles. We now wish to consider explicitly the detailed properties of these $G$-manifolds in the thermodynamic limit, but with $G$ finite. In particular, we wish to calculate the energy for states within the $G$-manifold, $E_G[K, M]$.

Since we are interested in the thermodynamic limit, we redefine our quantum numbers as densities:

$$\begin{aligned}
v &= M/L = \{M_1/L, \cdots, M_G/L\}, \\
\kappa &= 2\pi K/L = \{2\pi K_1/L, \cdots, 2\pi K_G/L\}, \\
n &= \{N_0/L, N_1/L, \cdots, N_G/L\}, \\
e_G[\kappa, v] &= (E_G[K, M | N] - E_0[N])/L.
\end{aligned} \qquad (26)$$

§7   The $G = 0$ subspace is the ground state; we now consider the $G = 1$ subspace, when a single internal mode is excited with quantum numbers $M_1 \equiv M$ and $K_1 = N_1 = N - N_0 \equiv K$. Thus, the space of quantum numbers is as shown in figure (2). We now examine the boundaries of this half-strip.

First, fix $M = 1$ and vary $K$ from 1 to $N-1$, or better yet $\kappa = 2\pi K/L$ from 0 to $2\pi d$ – the Brillouin zone. Such an excitation would usually be called a *phonon*, or a *hole* in the fermion picture, or perhaps a *fragmented condensate* in the boson picture. The energy is of order 1, written as

$$\varepsilon[\kappa | d] = E_{G=1}[K, M = 1 | N, L] - E_0[N, L]. \qquad (27)$$

and is given by solving the equation [4]

$$\varepsilon(k) = k^2/2 - \mu - \frac{1}{2\pi} \int_{-q}^{q} \theta'(k - k')\varepsilon(k')dk', \quad |k| \le q. \qquad (28)$$

The chemical potential $\mu$ is determined by the constraint that $\varepsilon(\pm q) = 0$. Likewise, the momentum is

$$P[\kappa | d] = P_{G=1}[K, M = 1 | N, L] = \kappa = 2\pi \int_{k}^{q} \rho(k')dk' \qquad (29)$$

where $\rho(k)$ is the solution to the integral equation

$$2\pi\rho(k) = 1 - \int_{-q}^{q} \theta'(k - k')\rho(k')dk', \quad |k| \le q. \qquad (30)$$

This is identical to eq. (14) for the ground state, in the center of mass frame, when $k_2 = q$, $k_1 = -q$. The density is again given by

$$d = \int_{-q}^{q} \rho(k)dk. \qquad (31)$$

This gives us our dispersion relation parametrically as

$$\Delta E = \varepsilon(k),$$

$$\Delta P = 2\pi \int_k^q \rho(k')dk', \tag{32}$$

$$v(k) = \frac{\varepsilon'(k)}{2\pi\rho(k)};$$

with $q \geq k \geq -q$. We remark that this is the ground state for fixed momentum. And to reemphasize – such an excitation is a single phonon in the ground state – a quantum fluid, rather than the classical lattice.

Second, fix $K = 1 = N_1$ so that $N_0 = N - 1$, and vary $M$, or better, $v = M/L$. Such an excitation will be called a *right-moving soliton*, or a *particle* in the fermion picture, or a *particle out of the condensate* in the boson picture. The energy is again of order 1, written as

$$\varepsilon_\sigma[v \mid d] = E_{G=1}[K = 1, M \mid N, L] - E_0[N, L]. \tag{33}$$

and is given by calculating the following expression using the previous solution $\varepsilon(k)$ to eq. (28):

$$\varepsilon_\sigma(k) \equiv k^2/2 - \mu - \frac{1}{2\pi}\int_{-q}^{q} \theta'(k-k')\varepsilon(k')dk', \quad k > q. \tag{34}$$

Likewise, the momentum is

$$P[v \mid d] = P_{G=1}[K = 1, M \mid N, L] = 2\pi \int_q^k \rho_\sigma(k')dk' = 2\pi v, \tag{35}$$

where $\rho_\sigma(k)$ is calculated from $\rho(k)$ by

$$2\pi\rho_\sigma(k) = 1 - \int_{-q}^{q} \theta'(k-k')\rho(k')dk', \quad k > q. \tag{36}$$

This gives us our dispersion relation parametrically as

$$\Delta E = \varepsilon_\sigma(k),$$

$$\Delta P = 2\pi v = 2\pi \int_q^k \rho_\sigma(k')dk', \tag{37}$$

$$v_\sigma(k) = \frac{\varepsilon_\sigma'(k)}{2\pi\rho_\sigma(k)};$$

with $k \geq q$. We note that $v(q) = v_\sigma(q) \equiv v_s$, the velocity of sound. Also, we identify the soliton velocity as the angular velocity $\omega = \frac{d\vartheta}{dt} = \frac{\partial H}{\partial A} = \frac{\partial \varepsilon_\sigma}{\partial v}$ for the mode $K = 1$. When the quantum number $M$ becomes large, we enter the semiclassical regime, allowing us to make a superposition of different $M$'s to fix the angle variable $\vartheta$, consistent with the uncertainty relation. In this way we identify the excitation with a classical soliton in a quantum fluid.

The third portion of the boundary $K = N - 1 = N_1$, $N_0 = 1$, is similarly evaluated to yield a *left-moving soliton*.

With finite numbers of these excitations, we can construct others, to order $1/L$ – the interaction energy. For instance, with $N \gg M > 1$ and fixed $K$, the energy and momentum of this coherent beam of phonons are simply $M$ times the energy and momentum of a single photon. Likewise, with $N \gg K > 1$ and fixed $M$, the energy and momentum of this *soliton train* of $K$ equally spaced solitons are simply $K$ times the energy and momentum of a single soliton of momentum $2\pi v$. Finally, we can add the energy and momentum of a finite number $G$ of different phonon beams and soliton trains to give the low-lying states (of order $1/L$) for different $H_G$. The conformal cone is a special case of this.

This takes care of the boundary of the state space for $G = 1$; we now look to the interior. Necessarily the values of $K$ and $M$ are large, and so we will be in the semi-classical regime. We will call such states solutions (of the equations of motion) of constant profile. In the asymptotic Bethe ansatz, for such states the asymptotic momenta distribute in two bands with band edges $k_4 > k_3 > k_2 > k_1$, and so the equations are

$$1 = 2\pi\rho(k) + \int_{k_1}^{k_2} \theta'(k-k')\rho(k')dk' + \int_{k_3}^{k_4} \theta'(k-k')\rho(k')dk', \tag{38}$$
$$k_2 > k > k_1 \quad \text{or} \quad k_4 > k > k_3.$$

It is also convenient to define the density of holes by

$$1 = 2\pi\rho_\sigma(k) + \int_{k_1}^{k_2} \theta'(k-k')\rho(k')dk' + \int_{k_3}^{k_4} \theta'(k-k')\rho(k')dk', \tag{39}$$

$$k_3 > k > k_2.$$

Then, we can calculate the physical quantities in terms of the band edges $k_4 > k_3 > k_2 > k_1$, using the following expressions:

$$\begin{aligned}
d &= \int_{k_1}^{k_2} \rho(k)dk + \int_{k_3}^{k_4} \rho(k)dk, \\
n_1 &\equiv \kappa/2\pi = \int_{k_1}^{k_2} \rho(k)dk, \\
m_1 &\equiv v = \int_{k_2}^{k_3} \rho_\sigma(k)dk, \\
P/L &= p = \hbar[\int_{k_1}^{k_2} + \int_{k_3}^{k_4}]k\rho(k)dk, \\
E/L &= e = \frac{\hbar^2}{2m}[\int_{k_1}^{k_2} + \int_{k_3}^{k_4}]k^2\rho(k)dk.
\end{aligned} \tag{40}$$

This allows us to parametrically determine the energy $e_1[\kappa,v\,|\,d,p]$ of the $G=1$ energy eigenstates; the parameters are the band edges $k_4 > k_3 > k_2 > k_1$.

However, we have agreed to work in the center of mass frame, so transforming, we obtain

$$e_1[\kappa,v\,|\,d] \equiv e_1[\kappa,v\,|\,d,0] = e_1[\kappa,v\,|\,d,p] - \frac{p^2}{2md}. \tag{41}$$

The interpretation is that in the center of mass frame, we have a beam with wave vector $\kappa$ and an intensity of $vL$ phonons; this beam has an energy density $e_1[\kappa,v\,|\,d]$ and momentum density $p[\kappa,v\,|\,d] = \hbar\kappa v$. We find the chemical potential for phonons to be

$$\frac{\partial e_1[\kappa,v\,|\,d]}{\partial v} \equiv \varepsilon[\kappa,v\,|\,d]. \tag{42}$$

and thus, in the limit of vanishing intensity, we recover $\varepsilon[\kappa,0\,|\,d] = \varepsilon[\kappa\,|\,d]$. However, since the action is given by $A/L = \hbar v$, we see that also

$$\frac{\partial e_1[\kappa,v\,|\,d]}{\partial v} = \hbar\omega[\kappa,v\,|\,d] = \varepsilon[\kappa,v\,|\,d]. \tag{43}$$

The quantum group velocity is

$$v = \frac{\partial e_1/\partial v}{\partial p/\partial v} = \frac{\omega}{\kappa}.$$

Similarly, viewing the beam as a soliton train with soliton density $\kappa/2\pi$, the chemical potential for solitons is $2\pi\partial e_1/\partial\kappa$, and the quantum group velocity is

$$v_\sigma = \frac{\partial e_1/\partial\kappa}{\partial p/\partial\kappa} = \frac{\partial e_1/\partial\kappa}{\hbar v}. \tag{44}$$

§8 We can understand the eigenstates in the interior of the phase plane by a hydrodynamic argument. We begin in the ground state with a uniform density $d=1/a$. Then we imagine 'particles' located on the average at positions

$$x_j = j/d = ja. \tag{45}$$

Note that this is the center of mass frame. We now consider another state with density $d(x,t)$ depending on both position and time. The average density remains $d$, and we stay in the center of mass frame. Now the 'particles' are located at $x_j(t)$, given by

$$j = \int_0^{x_j(t)} d(x,t)dx. \tag{46}$$

For states in the interior of the phase plane, the quantum numbers are large, and so we can make a wave packet minimizing the uncertainty relation, and thus effectively fixing both the action and angle variables. Let us consider the $G=1$ subspace. The density $d(x,t)$ is no longer translationally invariant, and so the time average must equal the space average. The $G=1$ subspace has a single angular velocity $\omega(\kappa)$, given by

$$\frac{\partial e_1[\kappa,v\,|\,d]}{\partial v} = \hbar\frac{\partial H}{\partial A} = \hbar\omega(\kappa). \tag{47}$$

We then conclude that the 'particle' coordinates are necessarily of the form

$$x_j(t) = [f(\omega(\kappa)t - j\kappa a) + j]a, \tag{48}$$

where $f(\vartheta)$ is a periodic function of $\vartheta$ with period $2\pi$. The angle variable is given by $\vartheta(t) = \omega(\kappa)t$, and we can recover the density as

$$d^{-1}(x,t) = [1 - \kappa a f'(\omega t - \kappa x)]a. \tag{49}$$

Thus, these states correspond to periodic density waves with constant profile, frequency $\omega(\kappa)$, wave vector $\kappa$, and a velocity given by the phase velocity $v = \omega(\kappa)/\kappa$, identical to the quantum group velocity. The periodic function also depends on the parameters $\kappa, v, d$. For a finite system, $\kappa$ must be discrete, in integer multiples of $2\pi/L$.

§9 As a simple example, let us take the inverse-square potential with potential.

$$U(r) = \frac{g}{r^2} \equiv \frac{\hbar^2 \lambda(\lambda-1)}{mr^2} \tag{50}$$

This has the very simple phase shift

$$\theta'(k) = 2\pi(\lambda-1)\delta(k), \tag{51}$$

and so,

$$\rho(k) = \frac{1}{2\pi\lambda}, \tag{52}$$
$$k_2 > k > k_1 \text{ or } k_4 > k > k_3.$$

The density of holes is

$$\rho_h(k) = \frac{1}{2\pi}, \tag{53}$$
$$k_3 > k > k_2.$$

The parameters are easily evaluated as

$$n_1 = \frac{1}{2\pi\lambda}[k_2 - k_1]$$

$$n_2 = \frac{1}{2\pi\lambda}[k_4 - k_3],$$

$$\kappa = \frac{1}{\lambda}[k_4 - k_3]'$$

$$v = \frac{1}{2\pi}[k_3 - k_2],$$

$$p = \frac{\hbar}{4\pi\lambda}[k_4^2 - k_3^2 + k_2^2 - k_1^2],$$

$$e = \frac{\hbar^2}{12\pi\lambda m}[k_4^3 - k_3^3 + k_2^3 - k_1^3]. \tag{54}$$

For this simple example we can explicitly invert to give the center of mass energy density as

$$e_1[\kappa, v \mid d] = \frac{\hbar^2}{m}[\pi^3 \lambda^3 d^3 L + 3\lambda\kappa v(2\pi d - \kappa)(2d\lambda + v)]. \tag{55}$$

Then, the first derivatives are

$$\frac{\partial e_1}{\partial v} = \frac{\hbar^2}{m} 6\lambda\kappa(2\pi d - \kappa)(d\lambda + v),$$

$$\frac{\partial e_1}{\partial \kappa} = \frac{\hbar^2}{m} 6\lambda v(\pi d - \kappa)(2d\lambda + v). \tag{56}$$

At the boundary, we recover

$$\varepsilon[\kappa] = \lim_{v \to 0} \frac{\partial e_1}{\partial v} = \frac{\hbar^2}{m} 6d\lambda^2 \kappa(2\pi d - \kappa),$$

$$\varepsilon_\sigma[\kappa] = \lim_{v \to 0} \frac{\partial e_1}{\partial \kappa} = \frac{\hbar^2}{m} 6\lambda v(\pi d - \kappa)(2d\lambda + v). \tag{57}$$

§10  More generally, for an energy eigenstate of genus $\Gamma = G + 1$, the asymptotic momenta distribute into $\Gamma$ bands with band edges $k_{2\Gamma} > \cdots > k_j > \cdots > k_1$, and so the asymptotic Bethe ansatz equations determining the density $\rho(k)$ of asymptotic momenta are

$$2\pi\rho(k) = 1 - \sum_{\alpha=1}^{\Gamma} \int_{k_{2\alpha-1}}^{k_{2\alpha}} \theta'(k - k')\rho(k')dk', \quad k_{2\alpha} > k > k_{2\alpha-1}, \tag{58}$$

with $k$ restricted to the bands

$$k_{2\alpha} > k > k_{2\alpha-1}, \ \alpha = 1, \cdots, \Gamma. \tag{59}$$

It is also useful to define the density of holes by

$$2\pi\rho_h(k) \equiv 1 - \sum_{\alpha=1}^{\Gamma} \int_{k_{2\alpha-1}}^{k_{2\alpha}} \theta'(k-k')\rho(k')dk', \tag{60}$$

with $k$ now restricted to the gaps

$$k_{2\alpha+1} > k > k_{2\alpha}, \ \alpha = 1, \cdots, \Gamma. \tag{61}$$

(We have extended the definition of the band edges by including $k_{2\Gamma+1} = +\infty > k_{2\Gamma} > \cdots > k_j > \cdots > k_1 > k_0 = -\infty$.)

After we have solved the asymptotic Bethe ansatz equations for the density of the asymptotic momenta, we then calculate physical properties of the energy eigenstates. We use the phonon scheme with a density $v_\alpha = M_\alpha / L$ of phonons of momentum $\hbar\kappa_\alpha = 2\pi\hbar K_\alpha / L$.

$$n_\alpha = N_\alpha / L = \int_{k_{2\alpha-1}}^{k_{2\alpha}} \rho(k)dk, \ \alpha = 1, \cdots, \Gamma;$$

$$\kappa_\alpha = 2\pi[d - \sum_{\beta=1}^{\alpha} n_\beta];$$

$$v_\alpha = M_\alpha / L = \int_{k_{2\alpha}}^{k_{2\alpha+1}} \rho_h(k)dk, \ \alpha = 1, \cdots, G = \Gamma - 1;$$

$$d = N / L = \sum_{\alpha=1}^{n} n_\alpha; \tag{62}$$

$$p = P / L = \hbar \sum_{\alpha=1}^{\Gamma} \int_{k_{2\alpha-1}}^{k_{2\alpha}} k\rho(k)dk;$$

$$e_G + e_0 = E / L = \frac{\hbar^2}{2m} \sum_{\alpha=1}^{\Gamma} \int_{k_{2\alpha-1}}^{k_{2\alpha}} k^2\rho(k)dk.$$

This allows us to eliminate the band edges, and determine the energy density as $e_G[\kappa, v \mid d, p]$, and finally, the center of mass energy as

$$e_G[\kappa,v\,|\,d] = e_G[\kappa,v\,|\,d,p] - \frac{p^2}{2md}. \tag{63}$$

The center of mass energy acts as a Hamiltonian for the collective excitations, and so

$$\frac{\partial e_G}{\partial v_\alpha} = \hbar \frac{\partial H}{\partial A_\alpha} = \hbar \omega_\alpha. \tag{64}$$

This is the chemical potential for adding a single phonon of momentum $\hbar \kappa_\alpha$.

§11   This paper is descriptive, consisting of assertions and conjectures, rather than proofs. How might one confirm, or at least support, these views? Examples are a possibility. In our original paper on the asymptotic Bethe ansatz [3] we recovered Toda's classic results for solitons [5] in the Toda lattice. We have also been able to recover Toda's results for the genus $\Gamma = 2$ cnoidal waves from the Bethe ansatz equations; this derivation will be presented in a later paper. More generally, the asymptotic Bethe ansatz has been shown to correctly describe the classical ground state (genus $\Gamma = 2$) for the integrable potentials eq. (2). [6,7] Presumably one could extend these arguments at least to genus $\Gamma = 2$, by using the Lax matrix [4], and thus recover the equivalent of Toda's Chapter 4 [5] for the more general potential.

In addition, we will examine the hydrodynamic description of the quantum fluid in a later paper, as well as the connection between our phonon picture, and the more familiar bose fluid picture, as presented for instance in Anderson [8].

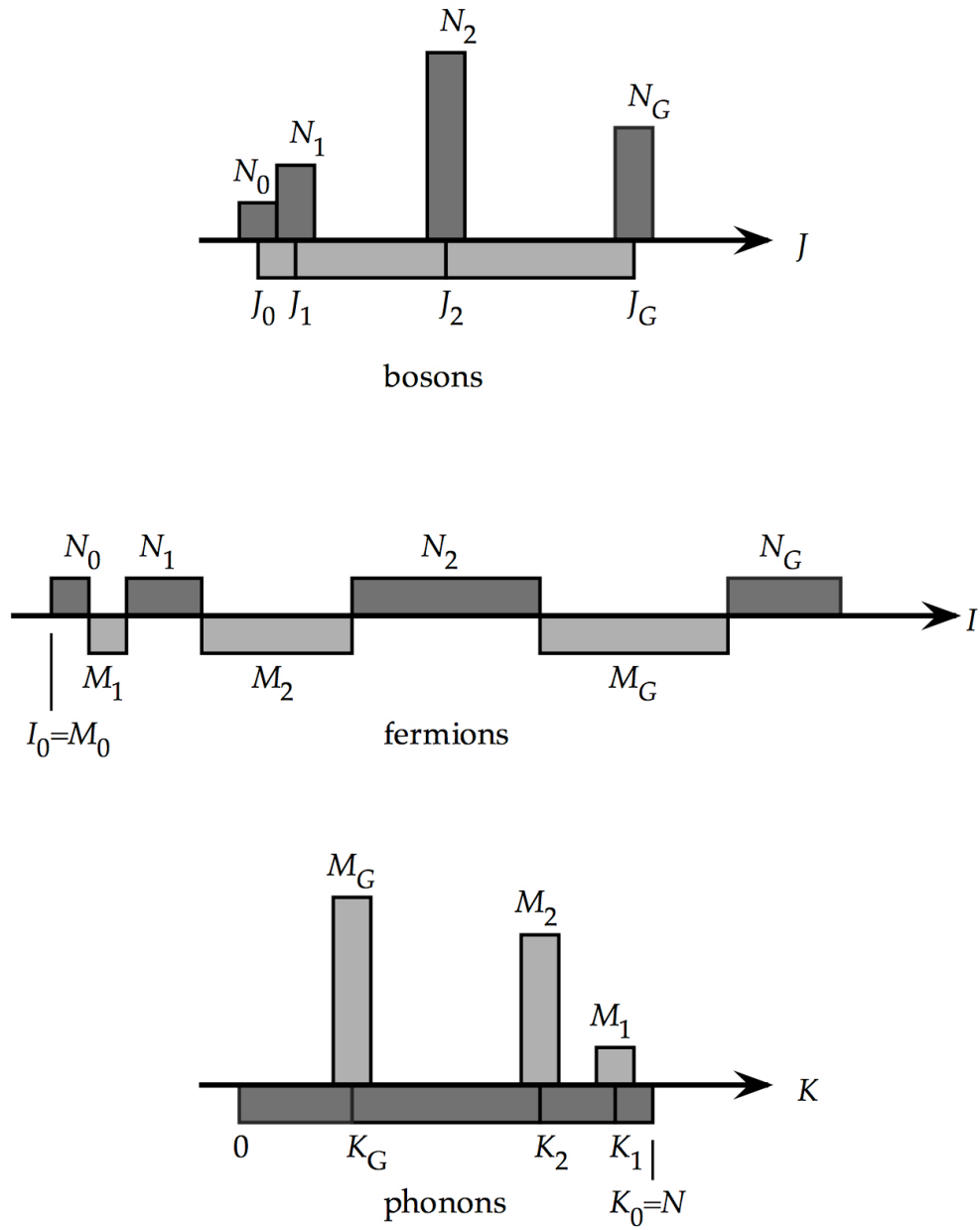

Figure 1  Illustration of the three schemes for a simple example with N=11.

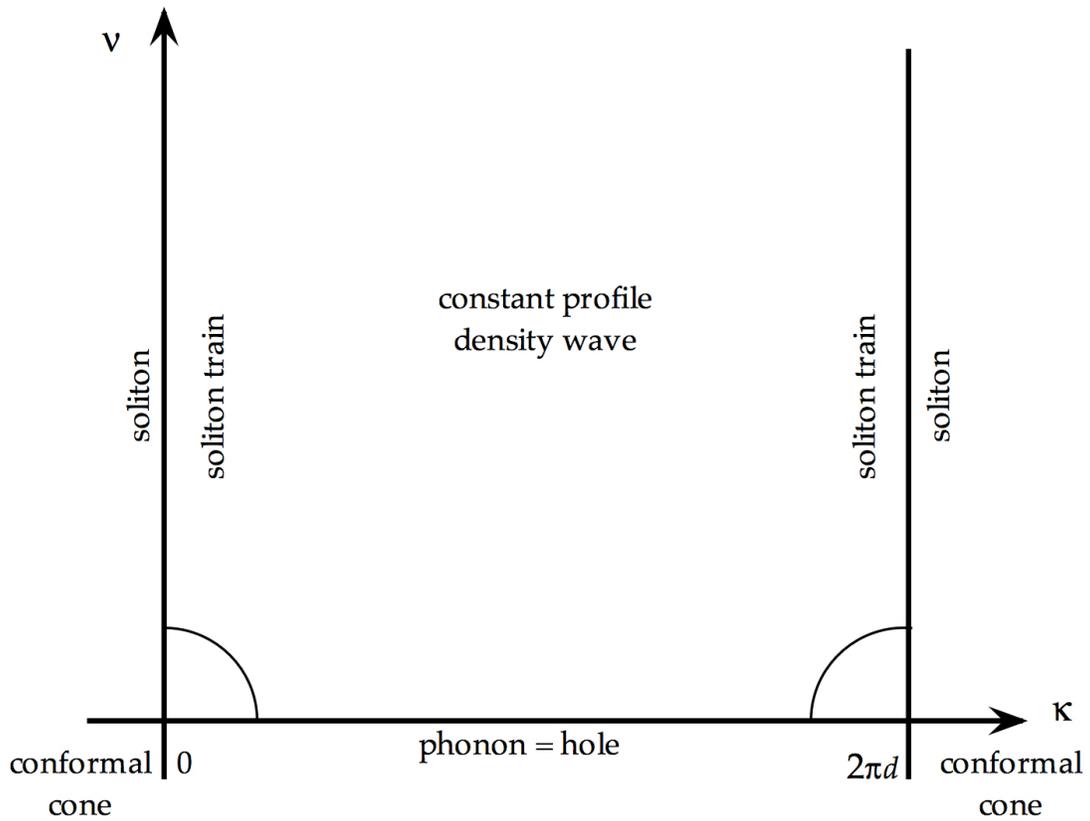

Figure 2   Phase plane for $G=1$.